\documentclass[12pt,preprint]{aastex}

\usepackage{color}

\newcommand{{\ka}}{${\rm K\alpha}$}

\shorttitle{SS~Cyg observations}
\shortauthors{Okada, Nakamura and Ishida}
\begin{document}

\title{{\it Chandra} HETG line spectroscopy
of the Non-magnetic Cataclysmic Variable SS~Cyg}

\author{{\sc Shunsaku OKADA}\altaffilmark{1}, {\sc Ryoko
NAKAMURA\altaffilmark{1}}, and {\sc Manabu ISHIDA}}
\affil{Institute of Space and Astronautical Science, 3-1-1, Yoshinodai,
Sagamihara, Kanagawa 229-8150, Japan}
\email{sokada@astro.isas.jaxa.jp, nakamura@astro.isas.jaxa.jp,
ishida@astro.isas.jaxa.jp}

\altaffiltext{1}{Also at Department of Physics, Tokyo Institute of
Technology, 2-12-1 Ohokayama, Meguro, Tokyo 152-8551, Japan}

\begin{abstract}

 We present {\it Chandra} HETG observations of SS~Cygni in quiescence
 and outburst. The spectra are characterized by He-like and H-like {\ka}
 emission lines from O to Fe, as well as L-shell emission lines from
 Fe. In quiescence, the spectra are dominated by the H-like {\ka} lines,
 whereas in outburst the He-like lines are as intense as the H-like
 lines. In outburst, the H-like {\ka} lines from O to Si are broad, with
 widths of 4--14$\ {\rm eV}$ in Gaussian $\sigma$ (1800--2300$\ {\rm km\
 s^{-1}}$). The large line widths, together with line profiles, indicate
 that the line-emitting plasma is associated with the Keplerian disk and
 still retains the azimuthal bulk motion. In quiescence, the emission
 lines are narrower, with a Gaussian $\sigma$ of 1--3$\ {\rm eV}$
 (420--620$\ {\rm km\ s^{-1}}$). A slightly larger velocity for lighter
 elements suggests that the lines in quiescence are emitted from an
 ionizing plasma at the entrance of the boundary layer, where the bulk
 motion of the optically thick accretion disk is converted into heat due
 to friction.  Using the line intensity ratio of He-like and H-like {\ka}
 lines for each element, we have also investigated the temperature
 distribution in the boundary layer both in quiescence and outburst.
 The distribution of SS~Cyg is found to be consistent with other dwarf
 novae investigated systematically with {\it ASCA} data.
\end{abstract}

\section{Introduction}

A dwarf nova is a close binary system consisting of a weakly magnetized
white dwarf and a low-mass companion filling its Roche lobe and is
characterized by frequent and large-amplitude optical outbursts.
Mass accretion from the companion occurs via an optically thick disk extending down close to the white dwarf, and a so-called
boundary layer is formed between the white dwarf and the innermost
edge of the accretion disk, where the accreting matter is braked
due to strong friction and finally settles onto the white dwarf.

After a long debate since the 1970's, it is now widely accepted that the
optical outburst arises due to a thermoviscous instability in an outer
region of the accretion disk.  In quiescence, matter from the companion
is accumulated in the disk because the hydrogen is neutral, and hence
the disk viscosity is low.  As matter accumulates and the temperature
exceeds $\sim$10$^4$~K, the viscosity increases due to hydrogen
ionization, and matter rushes onto the white dwarf at a rate exceeding
the supply rate from the companion.  This latter state is the optical
outburst
\citep{1984PASP...96....5S,1993ApJ...419..318C,1996PASP..108...39O,2001NewAR..45..449L}.
In response to the phase transition of the outer disk, it is predicted
that the physical state of the boundary layer changes significantly.  In
quiescence, the boundary layer is optically thin, because the mass
accretion rate is low and the radiative cooling is inefficient.  The
temperature is as high as $10^8$~K, and the boundary layer is a hard
X-ray emitter.  In outburst, on the other hand, the boundary layer can
remain optically thick just as for the outer region of the disk, because
of efficient radiative cooling.  Since its temperature is
$\sim\!10^5$~K, it emits mainly in the {\it EUV} band
\citep{1977MNRAS.178..195P,1979MNRAS.187..777P,1981AcA....31..267T,1985ApJ...292..535P}.
The first observational suggestion to support this phase transition was
obtained from a series of {\it European X-ray Observatory Satellite}
({\it EXOSAT}) observations \citep{1992MNRAS.257..633J} and was finally
established by \citet{2003MNRAS.345...49W}.

Although the phase transition of the boundary layer is better understood
than in the past, there still remain several unresolved issues.
Although the boundary layer is believed to be optically thick in
outburst, optically thin hard X-ray emission above 2~keV does not vanish
completely \citep{1994ApJ...436L..19N,2003MNRAS.345...49W}.  Unlike the
magnetic CVs in which the postshock accretion flow is essentially
one-dimensional, the temperature distribution of the boundary layer
plasma is not understood very well.  The temperature distribution is
probably tightly linked with the geometry of the boundary layer, for
which we have no more than a rough sketch.  In order to help resolve
these issues, we analyzed {\it Chandra} HETG data of SS~Cyg.  Since
SS~Cyg is the brightest dwarf nova in X-rays, it has been observed with
many X-ray astronomy satellites, such as {\it EXOSAT}, {\it Ginga}, the
{\it Rossi X-ray Timing Explorer} ({\it RXTE}), the {\it Advanced
Satellite for Cosmology and Astrophysics} ({\it ASCA}), {\it
Chandra}, and {\it Suzaku}
\citep{1992PASJ...44..537Y,1994ApJ...436L..19N,1997MNRAS.288..649D,2003MNRAS.345...49W,2004ApJ...610..422M,2005ASPC..330..355M,2007PThPS.169..178I}.
Since the {\it Chandra} HETG has an advantageous energy resolution, we
have attempted to approach these unresolved issues by means of
spectroscopy of He-like and H-like {\ka} emission lines.  We discuss the
temperature distribution and the geometrical structure of the boundary
layer based on our results.  In \S~\ref{sec:Observations} we introduce
the {\it Chandra} observations of SS~Cyg in outburst and quiescence and
present light curves and spectra. In \S~\ref{sec:DataAnalysis}, the
data analysis method and the results are presented. The results are
discussed in \S~\ref{sec:Discussion}, and the conclusions are summarized
in \S~\ref{sec:Conclusion}.

\section{Observations\label{sec:Observations}}
\subsection{{\it Chandra} data} 

The SS~Cyg data analyzed in this paper were retrieved from the {\it Chandra}
Data Archive\footnote{http://asc.harvard.edu/cda/} maintained by
the Smithonian Astrophysical Observatory.  The data in quiescence were obtained
between 10:28 UT on 2000 August 24 and 00:19 UT on August 25, with a
total good
time interval of 47.3 ks. The outburst data were obtained from 21:09 UT on 2000 September 14 to 14:15 UT on the following
day, with a total good exposure time of 59.5 ks. Both {\it Chandra}
observations employed the High Energy Transmission Grating (HETG)
\citep{2005PASP..117.1144C}, which allows us to carry out fine
spectroscopy with energy resolution better than CCD resolution by roughly an order
of magnitude. These data are analyzed also by
\citet{2003ApJ...586L..77M}, \citet{2005ASPC..330..355M}, and \citet{2006ApJ...642.1042R}. By referring to the light curve from the AAVSO home
page\footnote{http://www.aavso.org/}, we know that the visual magnitude
of SS~Cyg during the quiescence observation was in the usual range,
11.5--12.5. It then went into a normal outburst around September
10. Hence the {\it Chandra} outburst observation was performed $\sim$4--5
days after the onset of the outburst ($\sim$3 days after the peak with
$m_V \simeq 8.5$). The visual magnitude during the {\it Chandra}
observation was $m_V$ = 9.1--9.7.

\subsection{Light curves in quiescence and outburst}
\begin{figure}[htbp]
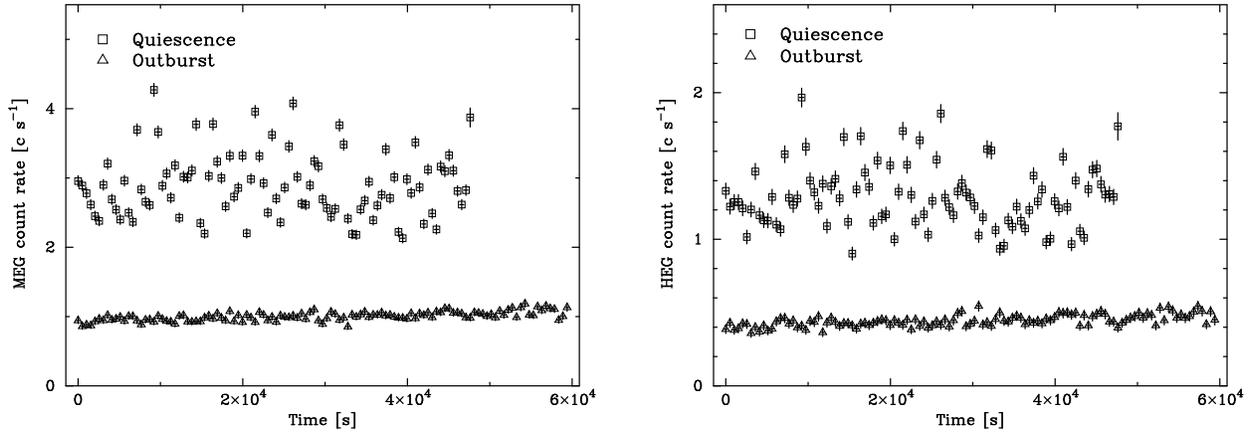

\begin{center}
\includegraphics[angle=270,scale=.32]{f1a.eps}
\hspace{0.5cm}
\includegraphics[angle=270,scale=.32]{f1b.eps}
\end{center}
\caption{HETG light curves of SS~Cyg in quiescence and outburst with a
 bin size of 512~s: (left) MEG light curves, (right) HEG light
 curves. The origin of the time axis coincides with the observation
 start time both for quiescence and outburst.\label{fig:LightCurve}}
\end{figure}

The HETG consists of two independent grating instruments: the Medium
Energy Grating (MEG), which covers the 0.4--5.0 keV band with an energy
resolution $ E/\Delta E$ of $\sim$200 at 2.0 keV, and the High Energy
Grating (HEG), which covers the 0.8--10.0 keV band with $E/\Delta E
\simeq$ 200 at 6.0~keV. In analyzing the data, we used
the CIAO software (Version 3.2) and CALDB (Version 3.1.0) provided by
the {\it Chandra} data center. Figure~\ref{fig:LightCurve} shows the
HETG light curves of SS~Cyg in quiescence and in outburst. These light
curves were made by summing up the positive and negative first-order
photons in the above energy bands with a 512~s bin width using the
thread {\tt dmextract}. Backgrounds are not
subtracted. The count rate of the MEG was ${\rm 2.90\pm 0.47\ c\
s^{-2}}$ and ${\rm 1.00\pm 0.07\ c\ s^{-1}}$ in quiescence and outburst,
respectively. That of the HEG was, on the other hand, ${\rm 1.29\pm
0.22\ c\ s^{-1}}$ and ${\rm 0.44\pm 0.04\ c\ s^{-1}}$ in quiescence and
outburst. The intensity of SS~Cyg in outburst is less than that in
quiescence by a factor of 3 on average and is less variable. The
weaker X-ray intensity during optical outburst is associated with the
optically thin-to-thick transition of the boundary layer when the mass
transfer rate in the outer disk exceeds a critical value
\citep{1977MNRAS.178..195P,1979MNRAS.187..777P,1981AcA....31..267T,1985ApJ...292..535P}.
Such events are detected by coordinated optical, EUV, and X-ray
observations \citep{2003MNRAS.345...49W}.

\subsection{HETG spectra in quiescence and outburst
\label{sec:HETGspecQandO}}
\begin{figure}[htbp]
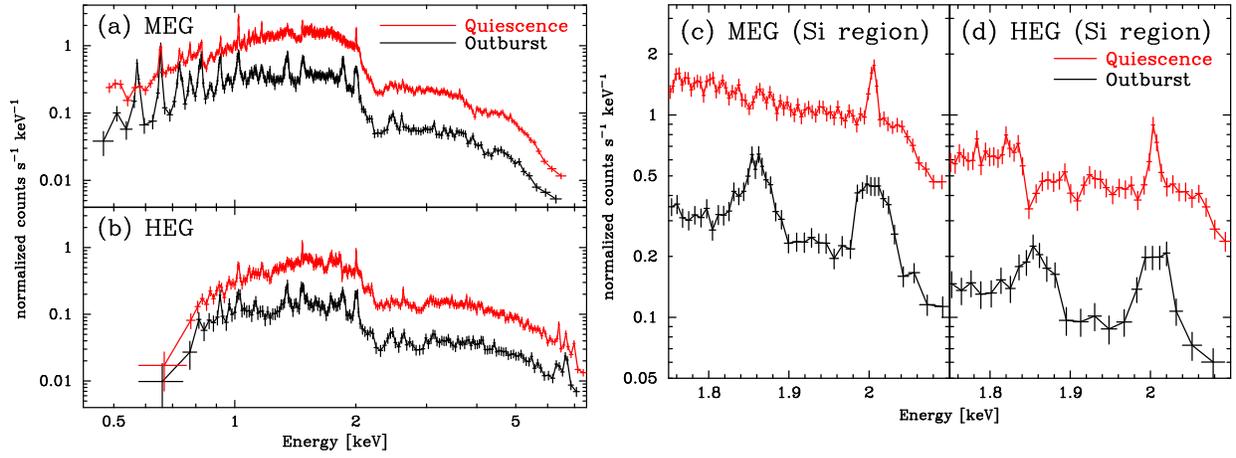

\begin{center}
\includegraphics[angle=270,scale=.32]{f2a.eps}
\includegraphics[angle=270,scale=.33]{f2b.eps}
\end{center}
\caption{Spectra of SS~Cyg in quiescence and outburst from (a) the MEG
 and (b) the HEG. Blow-up of (c) the MEG and (d) HEG spectra around the
 He-like and H-like Si {\ka} lines.\label{fig:Spectrum}}
\end{figure}

In Fig.~\ref{fig:Spectrum} we show spectra of SS~Cyg in quiescence and
outburst from the MEG and the HEG separately. 
The spectrum files (PHA files) were extracted from the level 1 data, and
the positive and negative orders were combined to enhance the
signal-to-noise ratio.  The MEG has a larger effective area in the lower
energy range and has clearly detected {\ka} emission lines from O, Ne,
Mg, and Si, while the HEG is advantageous at higher energies and has
detected {\ka} lines from Fe.  Simultaneous detection of the {\ka}
emission lines from these elements indicates the existence of optically
thin thermal plasma with a wide range of temperature distribution, both
in quiescence and outburst.  In quiescence, the spectra are dominated by
the H-like lines, whereas in outburst, the He-like lines are as strong as
the H-like lines. It is remarkable that, as demonstrated in
Fig.~\ref{fig:Spectrum}(c) and (d), the emission lines in outburst are
much broader than in quiescence. These characteristics
are also pointed out by \citet{2005ASPC..330..355M}. The profiles of
the H-like lines appear rectangular, rather than Gaussian-like.

\section{Data analysis\label{sec:DataAnalysis}}
As described in \S~1, we aim to elucidate the nature of the boundary
layer of SS~Cyg in terms of {\ka} emission lines from abundant metals.
We restrict ourselves to those from O up to Si, because the quantum
efficiency of both the MEG and HEG abruptly drops by a factor of $\sim$5
at the iridium M edge $\sim$2.1~keV (Fig.~\ref{fig:Spectrum}), above
which energy the line parameters are not very well constrained.  Since
the MEG has better sensitivity than the HEG in this energy band,
we concentrate on the MEG spectra, except for evaluation
of the H-like Si {\ka} line, for which we also utilize the HEG
data (\S~3.3 and \S~3.4).

\subsection{Continuum
\label{sec:continuum}}
In analyzing emission lines, we first attempted to evaluate the
continuum spectrum of the MEG in the 0.4--2.3 keV band.  We masked
energy bins, including the emission lines, and fitted the remaining
spectrum with a bremsstrahlung model undergoing a
uniform photoelectric absorption.  Prior to the fitting process, the
spectra were rebinned so that each energy bin includes at least 100
photons in quiescence and 50 photons in outburst, respectively.
The fits to the quiescence and outburst spectra were
marginally acceptable at the 90\% confidence level ($\chi^2$ [dof] of
452.5 [441] and 300.7 [241], respectively), with temperatures of ${\rm
14.5^{+4.8}_{-3.5}}$ and ${\rm 4.3^{+1.9}_{-0.7}\ keV}$ and with
$N_{\rm H}<0.3\times {\rm 10^{20}\ cm^{-2}}$ and ${\rm <1.1\times
10^{20}\ cm^{-2}}$, respectively. The fits were improved significantly
when we adopted a partially covering absorber with $\chi^2$ (d.o.f.) of
404.3 (440) and 259.1 (240).  The temperature of the
thermal bremsstrahlung in quiescence and outburst is ${\rm
4.1^{+1.4}_{-0.9}\ keV}$ and ${\rm 1.6^{+1.5}_{-0.9}\ keV}$, which is
covered with an absorber with $N_H=3.9^{+1.4}_{-0.9}\times 10^{19}\ {\rm
cm^{-2}}$ and $4.9^{+1.5}_{-0.9}\times 10^{19}\ {\rm cm^{-2}}$ by ${\rm
57^{+10}_{-9}}$\% and ${\rm 84^{+5}_{-7}}$\%, respectively. Below we adopt the latter as the continuum model.

\subsection{Emission lines
\label{sec:EvofLineParam}}
\begin{figure}[htbp]
\begin{center}
\includegraphics[angle=270,scale=.50]{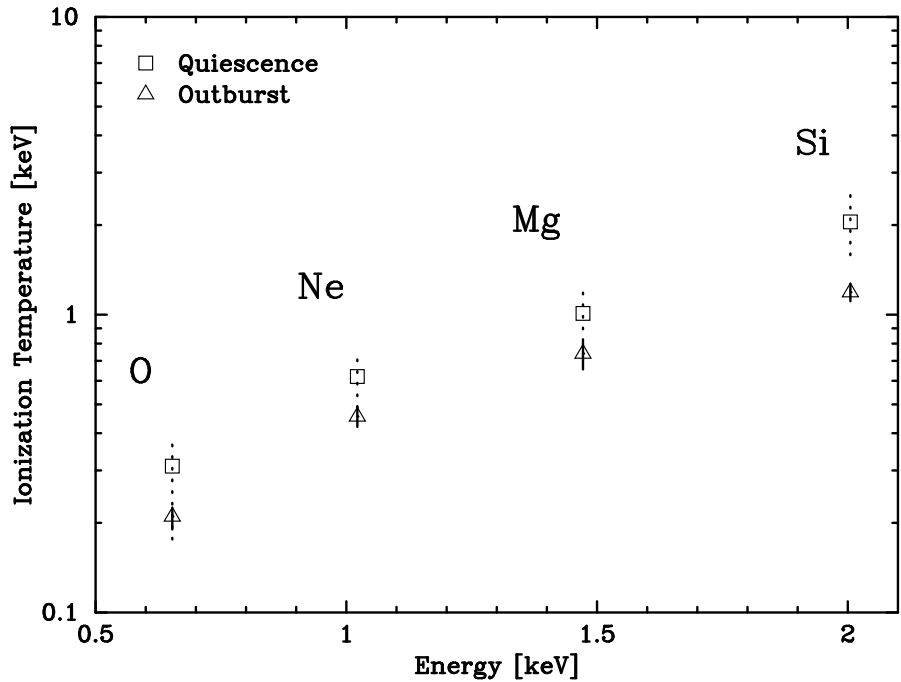}
\end{center}
\caption{The ionization temperatures obtained from the ratio of He-like
 and H-like {\ka} lines for each element.
\label{fig:IonizationTemperature}}
\end{figure}

Given the continuum model, we retrieved all the energy channels and
tried to evaluate the line parameters. We fixed all the continuum
parameters at the best-fit values in \S~\ref{sec:continuum}, except for
the normalization, and added Gaussians to represent the H-like and
He-like {\ka} lines of O, Ne, Mg, and Si.  In fitting these Gaussians to
the data, all the parameters were allowed to vary.
Table~\ref{table:IonizationTemperature} summarizes the best-fit
parameters of the {\ka} lines thus obtained in quiescence and outburst,
together with the ionization temperature obtained from the He-like and
H-like line intensity ratio, which are plotted in
Fig.~\ref{fig:IonizationTemperature}.

It is clear that the obtained ionization temperatures of all ions in
quiescence are systematically higher than those in outburst. They show a
wide range depending on the element, from 0.3 to 2.0 keV in quiescence
and from 0.2 to 1.2 keV in outburst. The emission lines seen in the
0.7--0.9$\ {\rm keV}$ bands are from L-shell Fe (mainly Ne-like),
indicating the existence of a plasma component with a temperature of
$\sim4\times 10^{6}\ {\rm K}$ \citep{1992ApJ...398..394A}. These facts
unambiguously indicate that the plasma in the boundary layer has a
significant temperature distribution. The differences in the line
intensity ratio distribution, as well as in the line profiles, probably
reflect a significant structural difference of the boundary layer
between quiescence and outburst.

\subsection{H-like line profile in outburst
\label{sec:HlineProfile}}
Unlike the He-like {\ka} line, which is composed of the resonance,
intercombination, and forbidden lines, as well as satellite lines, the
intrinsic structure of the H-like {\ka} line is much simpler,
consisting only of the following two resonance lines: 
$^2{\rm P}_{1/2}\rightarrow ^2\!{\rm S}_{1/2}$
and $^2{\rm P}_{3/2}\rightarrow ^2\!{\rm S}_{1/2}$.
Moreover, their energy separation is only 2.0~eV for Si ($E=2003.9$~eV
and 2005.9~eV, respectively)
and is even smaller for the lighter elements.
We can thus regard the H-like {\ka} lines of O, Ne, Mg, and Si
as consisting of a single component for the HETG resolution,
and their profile can be used to diagnose the geometry
and the dynamical motion of the line-emitting plasma.

\begin{figure}[htbp]
\begin{center}
 \includegraphics[width=.8\linewidth, height=.48\linewidth, angle=0]{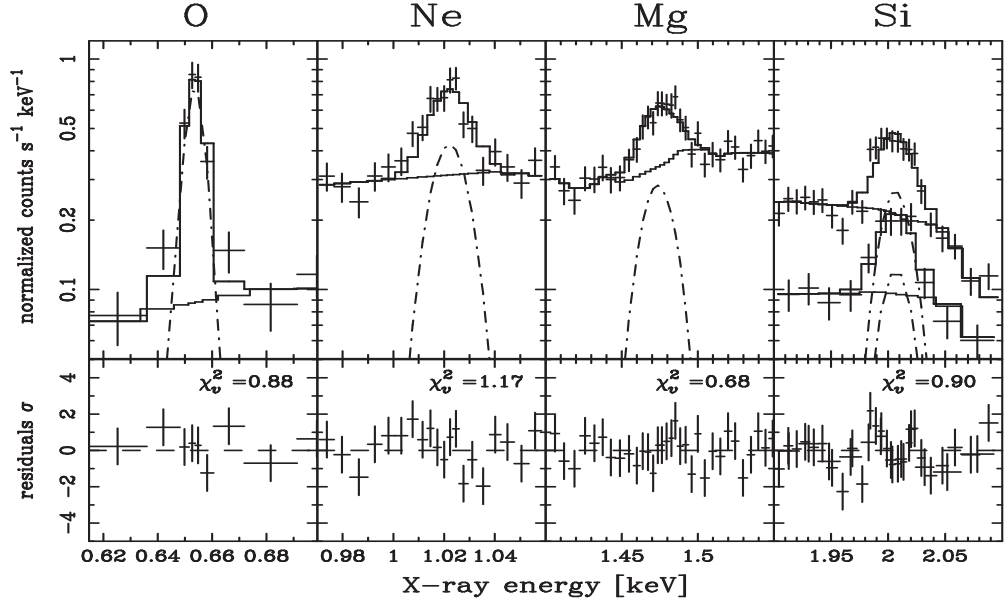}\\
\end{center}
\caption{Results of a Gaussian fit to the MEG spectra of the H-like
 {\ka} lines. The HEG spectra are also used
for the Si {\ka} line. The fits are acceptable at the 90\% confidence level.
 \label{Fig:singlegauss}}
\end{figure}

From Table~\ref{table:IonizationTemperature}, it is seen that all the
H-like {\ka} lines are significantly broader in outburst than in
quiescence (see also Fig.~\ref{fig:Spectrum}). Their widths are in the
range 4--14$\ {\rm eV}$ in Gaussian $\sigma$, and corresponding
velocities are 1800--2300$\ {\rm km\ s^{-1}}$. Their central energies
are consistent with those in the laboratory.  Their profiles, fitted
with a broad Gaussian, are shown in Fig.~\ref{Fig:singlegauss}.  We then
fixed $E_{\rm c}$ at the laboratory values, which are defined as the
emissivity-weighted means of the Ly$\alpha$ doublet.

The abscissa axis covers an energy range of 10\% of each line's
rest-frame energy.  The best-fit parameters fitted with the broad
Gaussian are summarized in table~\ref{table:singlegauss}.  Although all
the fits are already acceptable at the 90\% confidence level, the
profile seems to change from a Gaussian for lighter elements to a
somewhat rectangular structure for heavier elements.

The rectangular line profile is reminiscent
of a homogeneously emitting thin spherical shell
that is radially in- or outflowing \citep{2002A&A...392..991H}.
Of these, an outflow associated with SS~Cyg in outburst
is detected as the disk wind by the {\it Chandra} LETG
\citep{2004ApJ...610..422M}.
The wind is detected as blueshifted absorption lines
with a velocity of 2500$\ {\rm km\ s^{-1}}$ due to intermediately ionized
O through Fe, overlaid on a blackbody emission with a temperature of
$250 \pm 50$kK. Based on this fact, we infer that the observed H-like
emission lines would also have been blueshifted, because the plasma below
the accretion disk is invisible to us. We thus consider the radial
outflow picture unlikely.

Compared with the outflow considered above, a radial inflow is more
likely if the innermost part of the boundary layer becomes optically
thin, as in quiescence.
The plasma in the boundary layer then should take the form of a cooling
flow for matter settling down onto the white dwarf.
The temperature distribution demonstrated in
\S~\ref{sec:EvofLineParam} is interpreted as a radial stratification in
the cooling flow.
The plasma emitting any given line can be treated as a geometrically
thin spherical shell, since the emissivity of any {\ka} line is so
sensitive to the plasma temperature.

\begin{figure}[tbh]
\includegraphics[angle=270,scale=.45]{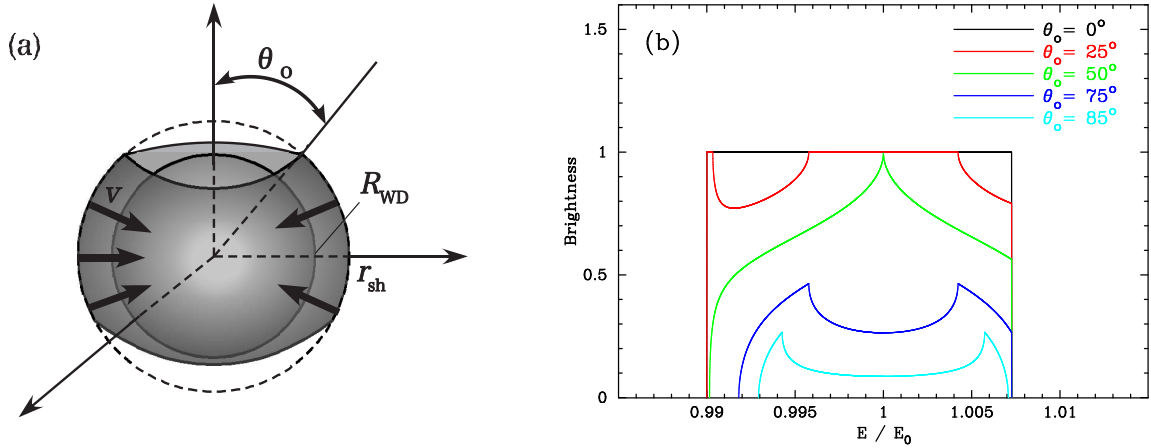} \hspace{1cm}
\includegraphics[angle=270,scale=.33]{f5b.eps}\\ 
\caption{(a) Schematic view of the stratified shell model, and (b) an
example of the line profiles. The adopted parameters are
$i=$40$^{\circ}$, $r_{\rm sh}/R_{\rm WD}=1.45$, and $v=3000\ {\rm km\
s^{-1}}$, and the line profiles are shown at opening angles $\theta_{\rm
o}=$ 0$^{\circ}$, 25$^{\circ}$, 50$^{\circ}$, 75$^{\circ}$, and
85$^{\circ}$.\label{fig:trapezmodel}}
\end{figure}

In order to see whether the radial inflow interpretation is acceptable, we
have constructed a model representing a profile of an emission line
emanating from a geometrically thin uniform shell
falling onto a white dwarf.
The free parameters of the model are the radial velocity $v$,
the radius of the shell $r_{\rm sh}$,
the opening polar angle of the shell $\theta_{\rm o}$,
the inclination angle of the orbital plane $i$, and the
central energy of the emission line $E_{\rm c}$ (see
Fig.~\ref{fig:trapezmodel}a).

The white dwarf always hides the most blueshifted part of the shell,
and the hidden fraction is determined solely by $r_{\rm sh}/R_{\rm WD}$,
where $R_{\rm WD}$ is the white dwarf radius.
In the case that the shell does not expand
above the orbital plane enough to form a complete sphere,
we introduce the polar opening angle $\theta_{\rm o}$ 
within which the plasma is absent. 
Some examples of the resultant line profiles are shown 
in Fig.~\ref{fig:trapezmodel}(b).
Hereafter we refer to this model as the stratified shell model.

\begin{figure}[htbp]
\begin{center}
\includegraphics[width=.8\linewidth, height=.83\linewidth, angle=0]{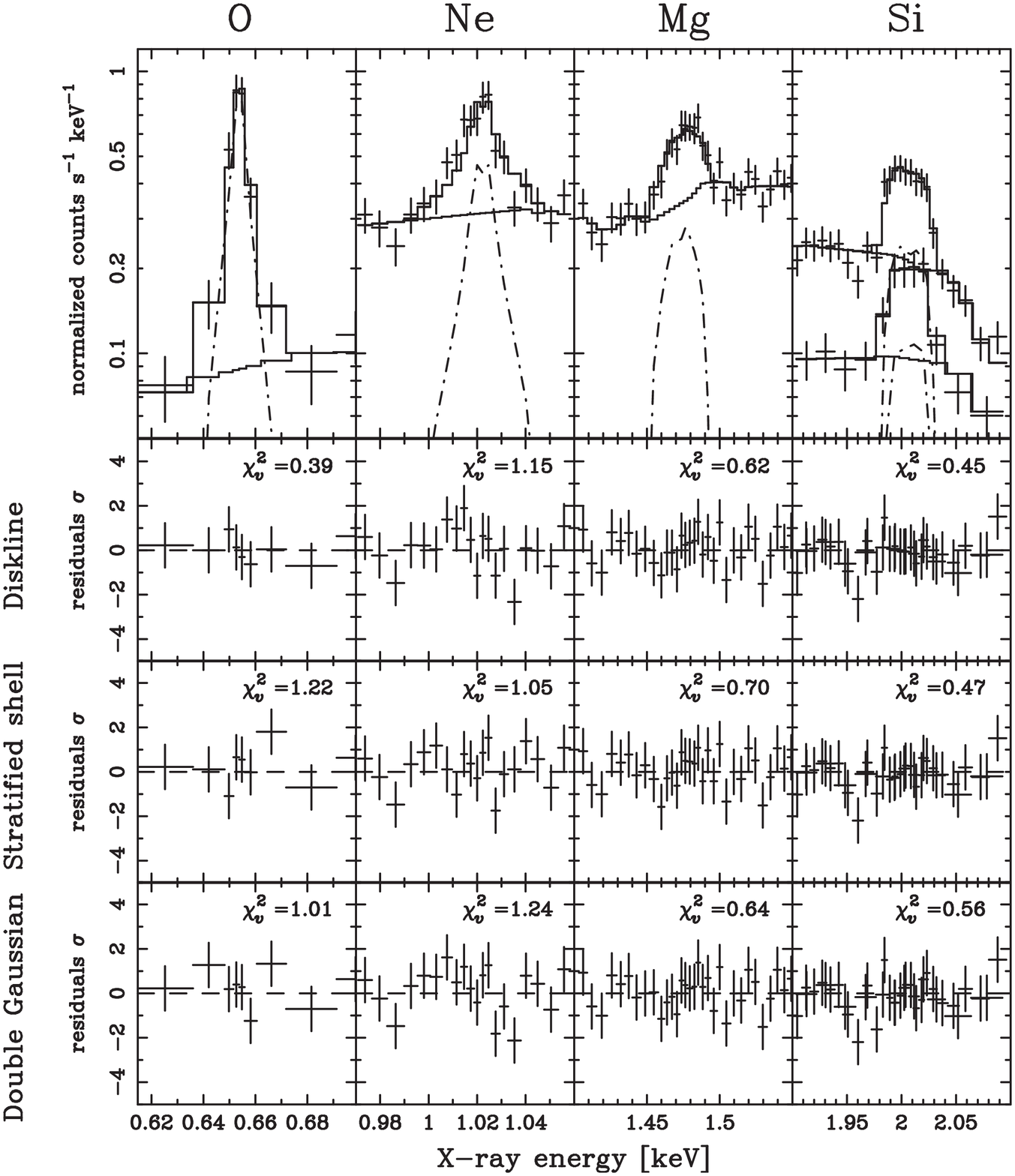}
\end{center}
\caption{Results of the fits of (top), the stratified shell model (middle)
 the diskline model, and (bottom) the double-Gaussian model to the
 H-like {\ka} lines.\label{Fig:manyfits}}
\end{figure}

In fitting the model to the H-like {\ka} lines, we fixed $i$ at
40$^\circ$, following the measurement of
\citet{1983PhDT........14S}. 
We also fixed $E_{\rm c}$ at the laboratory values. 
The best-fit parameters
of the lines with the stratified shell model are summarized in
table~\ref{table:trapez} and shown in Fig.~\ref{Fig:manyfits}.
As expected, the fitting to the H-like Si {\ka} line
is improved significantly, because the model can represent
the shoulders of the profile appropriately. The fitting to Mg is,
however, not improved significantly and that to Ne and O became poorer
than the Gaussian fit.
The radial velocity $v$ of all the elements is in the range
3400--4500~km~s$^{-1}$ and does not show any sign of slowing down
from the Si shell (higher $T$) to the O shell (lower $T$).

\begin{figure}[thb]
\begin{center}
\includegraphics[angle=0,scale=.33]{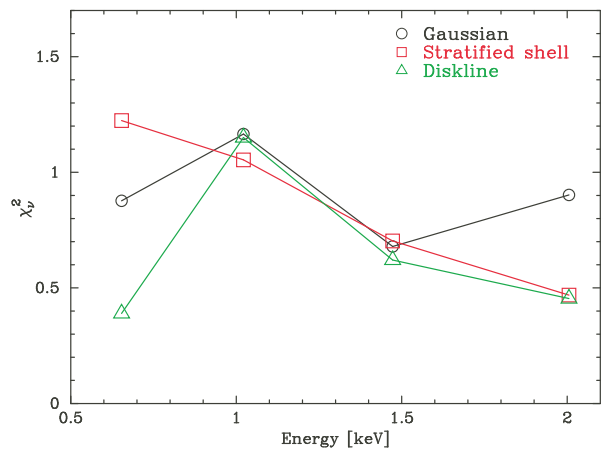}
\caption{$\chi^2_\nu$ values of the fits to the O through Si lines
with the single-Gaussian model, the stratified shell model, and the {\tt
 diskline} mode. \label{Fig:chisq}}
\end{center}
\end{figure}

Finally, we have tried the {\tt diskline} model in XSPEC
\citep{1996ASPC..101...17A} to see whether the profile
of the H-like {\ka} lines can be explained
by a bulk azimuthal motion of the plasma around the white dwarf.
The {\tt diskline} model is intended
to represent the emission line profile
from a geometrically thin Keplerian disk irradiated
by the central active galactic nucleus (AGN) via fluorescence, 
taking the gravitational redshift into account
\citep{1990Natur.344..747S, 1989MNRAS.238..729F}.
Hence, the parameters are the innermost radius of the disk
$R_{\rm in}$ in units of the Schwarzschild radius and the radial power $q$
of the emissivity $\varepsilon (r)$
in the form $\varepsilon \propto r^q$.
The results of the fits are
summarized in table~\ref{table:diskline} and displayed in
Fig.~\ref{Fig:manyfits}.
The resultant $\chi^2_\nu$ values of the fits to the O through Si lines
with the models introduced so far are shown in Fig.~\ref{Fig:chisq}.

The {\tt diskline} model can provide a comprehensive understanding
to a variety of the H-like {\ka} line profiles from O (Gaussian)
to Si (rectangular).
On the basis of this result, we conclude that the profiles of the H-like
{\ka} emission lines are best interpreted as due to azimuthal
bulk motion of the optically thin thermal plasma.

\subsection{Bulk motion of the plasma in outburst
\label{sec:PlaMotionOtb}}

As demonstrated in \S~\ref{sec:HlineProfile}, the nonrelativistic {\tt
diskline} model is successfully fitted to the H-like {\ka} lines from O
to Si in outburst.  This implies that the plasma has an azimuthal bulk
motion.  In order to evaluate this bulk motion, we hereafter fit the
profiles of the H-like {\ka} lines with two Gaussians that represent the
red- and blueshifted line components associated, respectively, with the
receding and approaching parts of the plasma.  The free parameters are
the average energy of the two line components $E_{\rm c}$, the energy
separation between them $E_{\rm sp}$, the energy widths $\sigma$, and
the normalizations.  Among these, $E_{\rm sp}$ is a measure of the
absolute radial location of the annulus from which the corresponding
H-like {\ka} line emanates, and $\sigma$ represents the geometrical
width of the annulus.  For simplicity, we set the $\sigma$ values, and
the normalizations of the red- and blueshifted components are
constrained to be common.

\begin{figure}
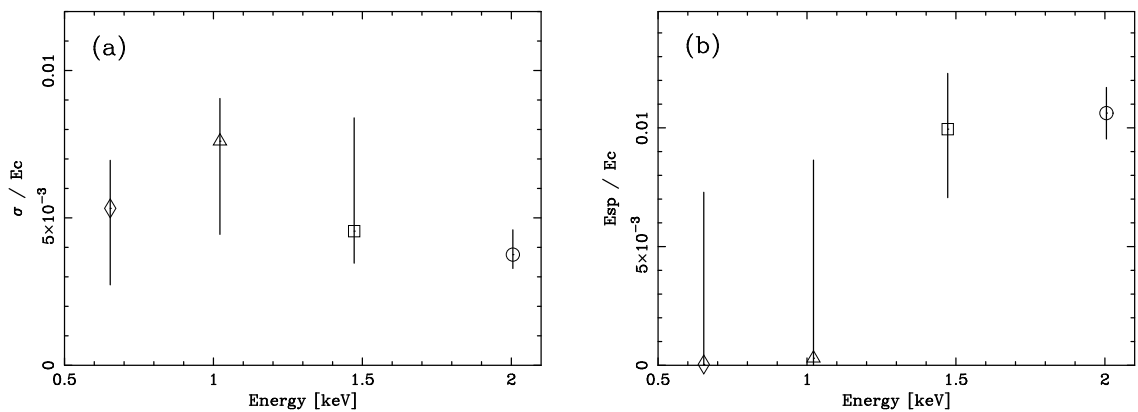

\begin{center}
\includegraphics[angle=270,scale=.33]{f8a.eps}
\hspace{5mm}
\includegraphics[angle=270,scale=.33]{f8b.eps}
\end{center}
\caption{(a) Resultant width $\sigma$ and (b) the separation energy
 $E_{\rm sp}$ of the two Gaussian line components in the double-Gaussian
 model, both normalized by the line central energy $E_c$.
 \label{Fig:doublegauss-results}}
\end{figure}

\begin{figure}[htbp]
\begin{center}
\includegraphics[angle=270,scale=.45]{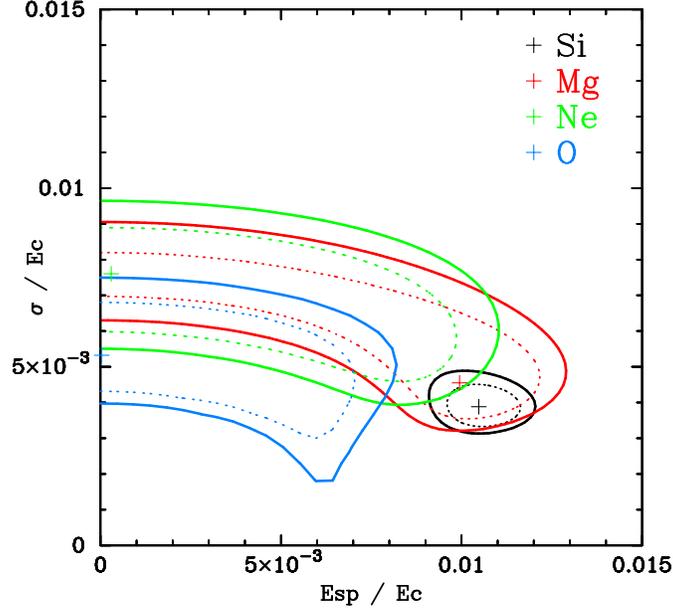}\\
\end{center}
\caption{The 90\% and 99\% confidence contours (dotted and continuous
 lines, respectively) between the line separation energy $E_{\rm sp}$
 and the line width $\sigma$ in the double-Gaussian model for the H-like
 {\ka} lines from O to Si. $E_{\rm sp}$ decreases from Si to O, and the
 90\% confidence contour of Si is isolated from that of O.
 \label{Fig:doublegauss-correlation}}
\end{figure}

The results of the fits are summarized in
Table~\ref{table:doublegauss} and shown in Fig.~\ref{Fig:manyfits} and
\ref{Fig:doublegauss-results}, with $\sigma$ and $E_{\rm sp}$ being
normalized by $E_{\rm c}$.
In addition, we draw confidence contours
between $E_{\rm sp}$ and $\sigma$
in Fig.~\ref{Fig:doublegauss-correlation},
since they may couple with each other.

From Fig.~\ref{Fig:doublegauss-results} and
\ref{Fig:doublegauss-correlation}, $E_{\rm sp}$ is found to be smaller for
lighter elements, and the 90\% confidence contours of O and Si are
separated, while $\sigma$ does not show any difference among the
elements.  In Table~\ref{table:doublegauss}, the rotational velocity $v$
calculated from $E_{\rm sp}$ with $v = (E_{\rm
sp}/2E_{\rm c})\,c$ is also listed.  The rotational velocity of Si is
$\sim$1500~km~s$^{-1}$, whereas that of O is only $\sim$10~km~s$^{-1}$.

\section{Discussion\label{sec:Discussion}}

\subsection{Geometry of line-emitting plasma in outburst}

In \S~\ref{sec:HlineProfile} and \S~\ref{sec:PlaMotionOtb}, we have
demonstrated that the profiles of H-like {\ka} lines from O to Si can be
explained with the azimuthal bulk motion model (the {\tt
diskline} model), rather than the radial inflow model (the stratified
shell model) on the basis of fit statistics, and we have obtained the
bulk motion parameters with the double-Gaussian model.
Besides the statistics, the high radial velocities of
$v=$~3400--4500~km~s$^{-1}$ (table~\ref{table:trapez}) are problematic
for the stratified shell model.
Since the ionization temperature of O in outburst is only $\sim$200~eV,
its thermal velocity is as small as
$\langle v_{\rm th}\rangle =\sqrt{3kT/m_{\rm O}} \simeq$350~km~s$^{-1}$.
Hence the plasma flow is highly supersonic, thereby forming
a standing shock wave whose temperature
$kT_{\rm s} = (3/16)\mu m_{\rm H} v^2$ is as high as 14--24~keV.
However, no such high temperature component is detected.
The {\it Suzaku} observations in outburst, for example,
constrain the maximum temperature of the plasma to be
$kT_{\rm max}{\rm =8.4\pm 1.0~keV}$ \citep{2007PThPS.169..178I}.

These results indicate that the bulk motion in the optically thin
thermal plasma in outburst is azimuthal rather than radial.
Although the errors are somewhat large, the lighter
elements are likely to have smaller $E_{\rm sp}$
(Fig.~\ref{Fig:doublegauss-results} and
\ref{Fig:doublegauss-correlation}). It is natural to expect that both
the temperature and the azimuthal bulk velocity are gradually reduced as
the plasma approaches the white dwarf, because it eventually settles
down onto the white dwarf. The double-Gaussian model is consistent with
this expectation in that the lighter elements, which have lower
ionization temperature, have smaller $E_{\rm sp}$.

Since the boundary layer is believed to be optically thick in outburst,
it has been unclear where the observed optically thin thermal X-ray
emission emanates.
Our analysis indicates that the optically thin thermal plasma also shows a
bulk azimuthal motion whose velocity decreases as the plasma cools and
approaches the white dwarf.
We infer that the optically thin
thermal plasma exists over the optically thick disk like an accretion
disk corona in low-mass X-ray binaries, as sketched in
\citet{1985ApJ...292..535P}.

\subsection{Line emission site in quiescence}
In \S~\ref{sec:HETGspecQandO} we demonstrated that the emission lines
in the quiescent spectra (Fig.~\ref{fig:Spectrum}) 
are significantly narrower than those in outburst.
Nevertheless, the lines still have finite widths
in the range 1--3~eV in Gaussian $\sigma$ for the H-like {\ka} lines
(Table~\ref{table:IonizationTemperature}).
They are as large as 1 part per $\sim\!\!10^3$ of the line energies
$E_{\rm c}$, and if they are attributed to thermal motion, then the
following holds:
\[
\frac{\sigma}{E_{\rm c}} \;=\;
 \frac{1}{c}\sqrt{\frac13 \langle v_{\rm th}^2 \rangle}
 \;=\; \frac{1}{c}\sqrt{\frac{kT}{m}}
\]
where ${\langle v_{\rm th}^2 \rangle}^{1/2}$ is the root mean square
of the thermal velocity. Using the atomic mass number $A$, we obtain
\begin{equation}
kT \;=\; mc^2\,\left( \frac{\sigma}{E_{\rm c}}\right)^2
 \;=\; 75\,\left( \frac{A}{20} \right)
 \,\left( \frac{\sigma}{\mbox{2 eV}}\right)^2
 \,\left( \frac{E_{\rm c}}{\mbox{1 keV}}\right)^{-2}\;\;\mbox{keV},
\end{equation}
which is 64~keV for O and 51~keV for Si.
These temperatures are incompatible with the detection
of the H-like {\ka} emission lines,
because they are emitted from a plasma with a temperature of 
$\lesssim 1$~keV \citep{1985A&AS...62..197M}.
Hence, the line broadening should originate from a bulk motion whose geometry is so symmetric that the line
Doppler shift is observed as a line broadening, not as a line shift.
One may consider an outflow such as is observed in
the EUV to soft X-ray band in outburst \citep{2004ApJ...610..422M}.
However, the outflow is probably located farther out of
the optically thin boundary layer, and most of the receding parts of the
plasma are hidden by the accretion disk.
In fact, the soft X-ray absorption lines originating from the outflow
in outburst are blueshifted \citep{2004ApJ...610..422M}.

We consider the observed {\ka} emission lines from O to Si as emanating
from a transition region from the optically thick accretion disk to the
optically thin boundary layer,
where the Keplerian motion is gradually converted into thermal motion
due to strong friction.
The average line-of-sight velocity dispersion
estimated from the observation is
\begin{equation}
 \langle v \rangle \;=\; \frac{\sigma}{E_{\rm c}}\,c
 \;=\; 6.0\times 10^2\,\left( \frac{\sigma}{\mbox{2 eV}}\right)\,
 \left( \frac{E_{\rm c}}{\mbox{1 keV}} \right)^{-1}\;\;{\rm km\;s^{-1}},
\label{eq:ave_vel}
\end{equation}
which is 620$^{+130}_{-170}$,
530$^{+100}_{-120}$, $460^{+220}_{-120}$,
and 420$\pm 140$~km~s$^{-1}$
for O, Ne, Mg, and Si, respectively.
This is roughly one-tenth of the Keplerian velocity on the surface of
the 1.2$M_\odot$ white dwarf ($\simeq$6300~km~s$^{-1}$).
Even considering the inclination angle
of 40$^\circ$
and the line-of-sight average of the azimuthal velocity,
$(2/\pi)\sin 40^\circ$,
we obtain the maximum possible line-of-sight velocity to be
$\langle v \rangle_{\rm max} = $2600~km~s$^{-1}$, 
which can cover the average velocity eq.~(\ref{eq:ave_vel})
evaluated from the observed line width.

As in outburst, the line width in quiescence can be attributed 
to azimuthal motion. Unlike in outburst, however, where the emission
lines can be interpreted as being emitted from a cooling plasma, we
believe those in quiescence are created in the heating process at the
entrance to the optically thin boundary layer.  In fact, although the
statistical significance is not very high, the bulk velocity in
quiescence estimated from the H-like {\ka} line widths seems to decrease
from O to Si.  It is likely that the plasma heating results
from the conversion of the energy of bulk motion.

\subsection{Temperature distribution }
\begin{figure}[htbp]
\begin{center}
\includegraphics[angle=0,scale=.33]{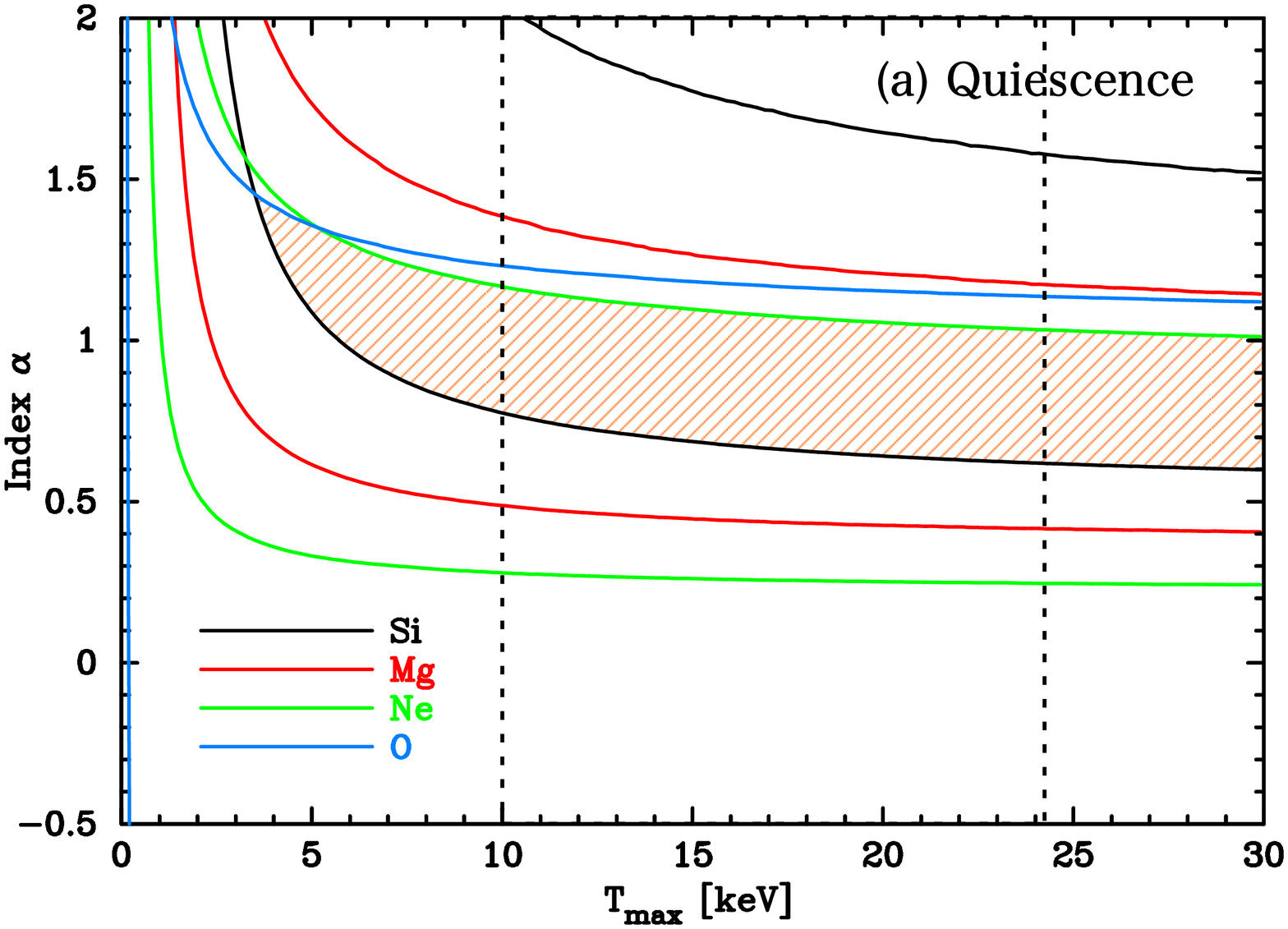}
\includegraphics[angle=0,scale=.33]{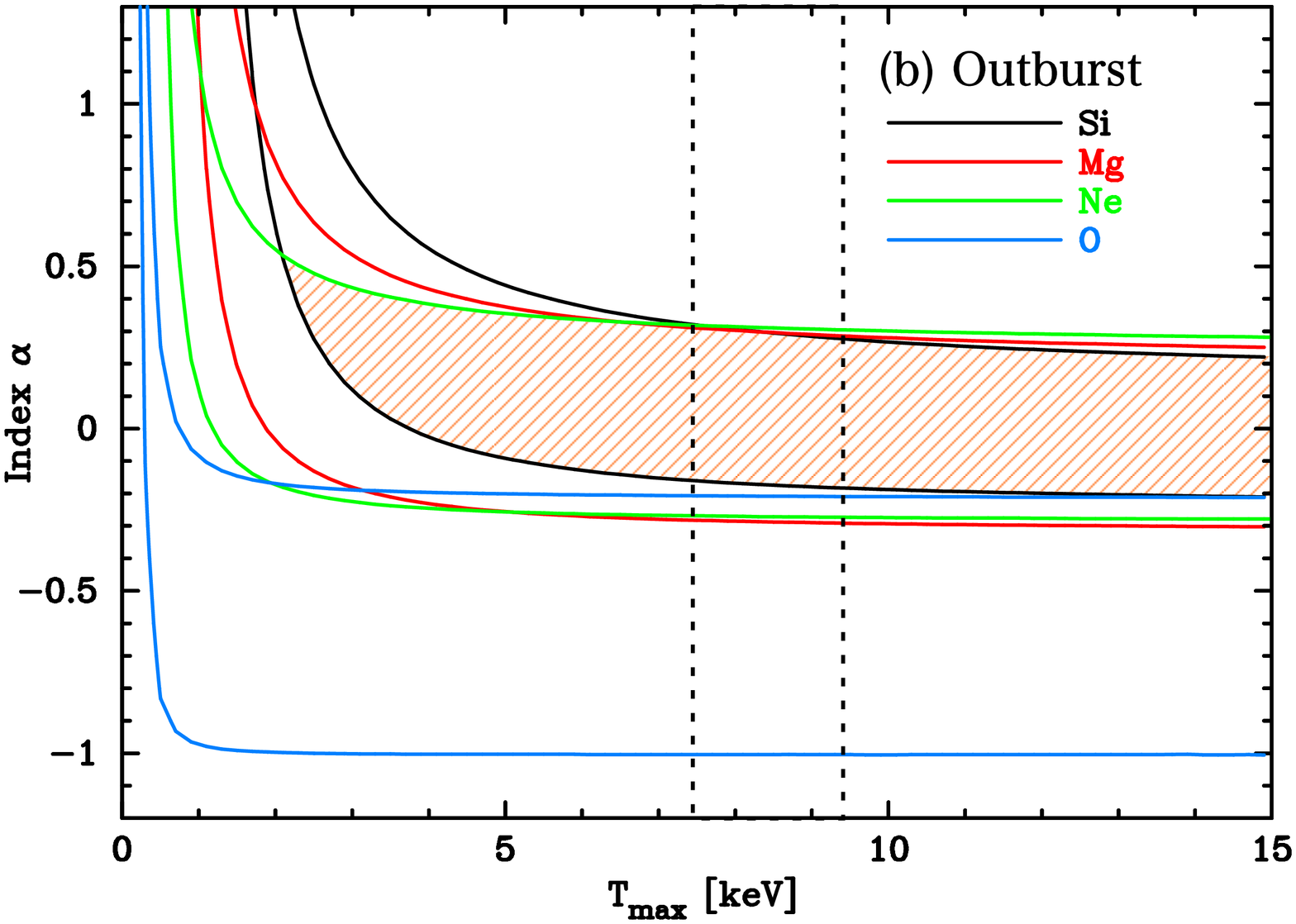}
\end{center}
\caption{Areas allowed by the observed line intensity ratio of O, Ne,
 Mg, and Si in the plane of the two {\tt cemekl} parameters $\alpha$ and
 $T_{max}$. The allowed areas are defined with the 90\% confidence
 contours. Left and right panels are those for quiescence and outburst,
 respectively. Vertical dotted lines at $15.4^{+8.8}_{-5.4}$ and $8.4\pm
 1.0$~keV are the values of $T_{\rm max}$ from
 \citet{2007PThPS.169..178I} and \citet{1997MNRAS.288..649D},
 respectively.  \label{Fig:lineratio-coutour}}
\end{figure}

In this section, we attempt to impose a constraint on the temperature
distribution of the plasma in outburst and quiescence in terms of the
intensity ratio of the He-like and H-like {\ka} emission lines.  We
simply assume that the differential emission measure $d(E\!M)$ is
proportional to the power of the plasma temperature.  In the following
analysis, we adopt the spectral model {\tt cemekl}, in which the
temperature distribution is taken into account as $d(E\!M)\propto
(T/T_{\rm max})^\alpha d(\log T)$ $\propto (T/T_{\rm
max})^{\alpha-1}dT$, where $T_{\rm max}$ is the maximum temperature of
the plasma.  Hence, the exponent $\alpha$ obtained by a fit with this model
is larger than the true weighting factor by 1
\citep{1997MNRAS.288..649D}.  In the framework of this model, the line
intensity ratio is uniquely determined once $\alpha$ and $T_{\rm max}$
are given.  Hence, we can conversely give a constraint on these two
parameters with the line intensity ratios of O, Ne, Mg, and Si that we have
obtained from the HETG data.  The allowed parameter regions delineated
by the 90\% errors of the intensity ratios
(Table~\ref{table:IonizationTemperature}) are shown in
Fig.~\ref{Fig:lineratio-coutour} with a different color for each element.

For all the elements, the contours become vertical, and $\alpha$
diverges as $T_{\rm max}$ decreases and approaches
a certain value.
This limiting $T_{\rm max}$ gives the temperature of the single-phase
plasma that can reproduce the observed intensity ratio.
On the contrary, the contour becomes horizontal
in the high-$T_{\rm max}$ limit.
Since the emissivity of atomic lines peaks at a finite temperature,
the line ratio becomes insensitive to $T_{\rm max}$
and is determined solely by $\alpha$ if $T_{\rm max}$
becomes too high compared with the emissivity peak temperature.
As shown in Fig.~\ref{Fig:lineratio-coutour},
there is a parameter region that can be shared
by all elements (except for O in outburst).
Taking into account the constraint $T_{\rm max} = 8.4\pm 1.0$~keV
for the outburst obtained from the {\it Suzaku} observation
\citep{2007PThPS.169..178I}, the common range of $\alpha$ 
is $-0.2$ to 0.3 in outburst.
On the other hand, the maximum temperature is $15.4^{+8.8}_{-5.4}$~keV
\citep{1997MNRAS.288..649D} in quiescence,
which is significantly higher than in outburst,
and the common range of $\alpha$ is found to be 0.6--1.2
in quiescence.
This significant difference is probably the result
of the geometrical difference of the optically thin boundary layer
between quiescence and outburst (for example,
see \citet{1985ApJ...292..535P}).

We remark that only the index $\alpha$ of O in outburst is not common with
the other elements. 
This is not unreasonable, in view of the shape of the cooling curve of
an optically thin thermal plasma.
Unlike the temperature region $T \gtrsim 2\times 10^7$~K,
the slope of the plasma cooling curve shows a complicated change as a
function of the temperature in the range $T \lesssim 10^7$~K
\citep{1993ApJ...418L..25G}.
Since $\alpha$ of a given element reflects the slope of the cooling
curve around the temperature at which the line emissivity of the
element shows the peak,
it is possible in principle that $\alpha$ is different for elements
whose emissivity peak appears in the range $T \lesssim 10^7$~K.

The index of the temperature distribution $\alpha$ was systematically
examined by \citet{2005MNRAS.357..626B} using the complete set of the
{\it ASCA} nonmagnetic CV data.
They found that $\alpha$ for the dwarf novae 
is in the range from 0.7 to 1.8 in quiescence (with a few exceptions)
and $-$0.1 to 0.3 in outburst.
The $\alpha$-values obtained from our analysis on SS~Cyg in quiescence
and outburst are both consistent with their results.
\citet{1997MNRAS.288..649D} carried out the same analysis using the
outburst and quiescence data of {\it Ginga} and {\it ASCA}.
Although $\alpha$ in the quiescence {\it Ginga} data is not
constrained very well ($2.9^{+\infty}_{-1.4}$),
they derived $\alpha = 0.43^{+0.41}_{-0.38}$ and $0.46^{+0.06}_{-0.05}$
for the normal and anomalous outbursts from the {\it Ginga} and {\it ASCA}
data, respectively.
The $\alpha$ of the anomalous outburst is outside the range of
\citet{2005MNRAS.357..626B} and that of ours.
The exponent $\alpha$ may possibly be different between the two types of
the outburst and may depend on the intensity of the reflection
component included in their modeling.

The temperature distribution in the postshock accretion column of a
magnetic cataclysmic variables (mCVs) can be approximated as $T(z)
\propto z^{2/5}$, where $z$ is the height from the white
dwarf surface if the cooling mechanism is dominated by free-free
emission \citep{2002apa..book.....F}.  In this case, the {\tt cevmekl}
power $\alpha \simeq +0.5$ \citep{1994MNRAS.266..367I},
which lies between that in the quiescence and the
outburst obtained from this work.

\section{Conclusion
\label{sec:Conclusion}}
We have investigated the optically thin boundary layer of SS~Cyg both in
quiescence and in outburst by carrying out high resolution line spectroscopy
with the {\it Chandra} HETG.
The spectra contain He-like and H-like {\ka} emission lines from O
to Fe, as well as L-shell emission lines from Fe,
both in quiescence and outburst.
However, the relative intensity of the He-like and H-like emission lines
of a given element is significantly different; in outburst they are
nearly equal, whereas in quiescence the H-like lines are much more intense
than the He-like line.
The ionization temperatures evaluated from the line
intensity ratio of O, Mg, Ne, and Si are in the range 0.3--2.0~keV and
0.2--1.2~keV in quiescence and outburst, respectively.

The other remarkable difference of the line profiles between quiescence
and outburst is in the line width.  The H-like lines from O to Si in
outburst are much broader, with widths of 4--14~eV in
Gaussian $\sigma$ (1800--2300~km~s$^{-1}$), and especially that of Si has
a rectangular profile, rather than a Gaussian profile.  Among a few
possibilities, such as a homogeneously emitting thin spherical shell
that is radially in- or outflowing, an azimuthal bulk motion, which
can be mimicked by the {\tt diskline} model, most naturally explains the
observed line profile.  The boundary layer is, however, believed to be
optically thick all the way down to the white dwarf surface.  We thus
infer that the line-emitting plasma is located above the optically thick
disk, like an accretion disk corona commonly seen in low-mass X-ray
binaries.  Since lighter elements tend to have a smaller bulk velocity,
the observed line-emitting plasma in outburst is a cooling plasma.

Although less remarkable, the emission lines from O to Si in quiescence
also have a finite line width with a Gaussian $\sigma$ of 1--3~eV
(420--620~km~s$^{-1}$).  It is impossible to attribute the width to a
thermal motion, because the resultant temperature becomes as high as
$\gtrsim$50~keV.  Since no red- or blueshift is detected for the line
central energies, the radial in- or outflow is unlikely.  The most
natural origin of the bulk velocity is an azimuthal motion.  A slightly
larger velocity for lighter elements, contrary to outburst, suggests
that the lines in quiescence are emitted in the heating process at the
entrance to the boundary layer, where the bulk motion of the optically
thick accretion disk is converted to thermal energy due to friction.

We have also investigated the temperature distribution in the boundary
layer within a framework of the differential emission measure having a
power-law distribution of the temperature as $(T/T_{\rm max})^\alpha$.
From the line intensity ratio,
the index $\alpha$ is constrained in the range from $-$0.2 to 0.3 in
outburst and from 0.6 to 1.2 in quiescence. These values are
consistent with those obtained from the previous systematic work
of \citet{2005MNRAS.357..626B}.

%
%
%
%
\begin{deluxetable}{lcccccc}
\tablecaption{Best-fit parameters of a Gaussian to the {\ka} lines and
 resultant ionization temperatures \label{table:IonizationTemperature}}
\tablewidth{0pt}
\tabletypesize{\small}
\tablehead{  & \colhead{$E_{\rm c}$} & \colhead{$\sigma$}
 & \colhead{$I$} & \colhead{EW} & \colhead{$I_{\rm H}/I_{\rm He}$} 
 & \colhead{$kT_i$} \\
 \colhead{Ion} & \colhead{[keV]} & \colhead{[eV]}
 & \colhead{[${\rm 10^{-4} ph.\;cm^{-2}s^{-1}}$]} & \colhead{[eV]} 
 & \colhead{}
 & \colhead{[keV]}
}
\startdata
\multicolumn{7}{c}{Quiescence} \\
\tableline
Si$\;${\tiny XIII} & ${\rm 1.8649^{+0.0030}_{-0.0058}}$
 & ${\rm 6.16^{+5.55}_{-3.16} }$ & ${\rm 0.55^{+0.26}_{-0.27}}$
 & ${\rm 2.9}$ & ...  & ... \\
Si$\;${\tiny XIV} & ${\rm 2.0052^{+0.0005}_{-0.0012}}$
 & ${\rm 2.80^{+0.94}_{-0.95} }$ & ${\rm 1.53^{+0.24}_{-0.25}}$
 & ${\rm 9.1}$ & ${\rm 2.79^{+1.44}_{-1.37}}$ & ${\rm 2.05^{+0.465}_{-0.53}}$ \\
Mg$\;${\tiny XI} & ${\rm 1.3521^{+0.0024}_{-0.0020}}$
 & ${\rm 4.17^{+3.83}_{-1.76} }$ & ${\rm 0.87^{+0.26}_{-0.40}}$
 & ${\rm 2.6}$ & ... & ... \\
Mg$\;${\tiny XII} & ${\rm 1.4720^{+0.0073}_{-0.0059}}$
 & ${\rm 2.25^{+1.06}_{-0.61} }$ & ${\rm 1.54^{+0.35}_{-0.26}}$
 & ${\rm 5.2}$ & ${\rm 1.78^{+0.90}_{-0.61}}$ &  ${\rm 1.01^{+0.19}_{-0.15}}$ \\
Ne$\;${\tiny IX} & ${\rm 0.9210^{+0.0017}_{-0.0009}}$
 & ${\rm 2.15^{+1.21}_{-0.95} }$ & ${\rm 2.23^{+1.03}_{-0.76}}$
 & ${\rm 3.5}$ & ... & ... \\
Ne$\;${\tiny X} & ${\rm 1.0213^{+0.0411}_{-0.0003}}$
 & ${\rm 1.81^{+0.33}_{-0.41} }$ & ${\rm 4.68^{+0.60}_{-0.76}}$
 & ${\rm 8.7}$ & ${\rm 2.10^{+0.76}_{-1.03}}$ & ${\rm 0.62^{+0.08}_{-0.13}}$ \\
O$\;${\tiny VII} & ${\rm 0.5702^{+0.0399}_{-0.0399}}$
 & ${\rm 3.26^{+9.34}_{-1.63} }$ & ${\rm 7.27^{+5.28}_{-6.34}}$
 & ${\rm 6.5}$ & ... & ... \\
O$\;${\tiny VIII} & ${\rm 0.6533^{+0.0004}_{-0.0003}}$
 & ${\rm 1.35^{+0.29}_{-0.36} }$ & ${\rm 13.78^{+2.27}_{-2.47}}$
 & ${\rm 14.9}$ & ${\rm 1.90^{+1.42}_{-1.68}}$ & ${\rm
 0.30^{+0.07}_{-0.15}}$ \\
\tableline
\multicolumn{7}{c}{Outburst} \\
\tableline
Si$\;${\tiny XIII} & ${\rm 1.8603^{+0.0011}_{-0.0030}}$
 & ${\rm 12.50^{+2.49}_{-1.00} }$ & ${\rm 4.62^{+0.71}_{-0.39}}$
 & ${\rm 46.4}$ & ... & ... \\
Si$\;${\tiny XIV} & ${\rm 2.0055^{+0.0023}_{-0.0023}}$
 & ${\rm 13.88^{+2.04}_{-1.32} }$ & ${\rm 3.73^{+0.34}_{-0.61}}$
 & ${\rm 45.4}$ & ${\rm 0.81^{+0.10}_{-0.18}}$ & ${\rm 1.22^{+0.053}_{-0.104}}$ \\
Mg$\;${\tiny XI} & ${\rm 1.3494^{+0.0010}_{-0.0034}}$
 & ${\rm 9.95^{+1.48}_{-1.66} }$ & ${\rm 5.52^{+0.57}_{-1.13}}$
 & ${\rm 25.9}$ & ... & ... \\
Mg$\;${\tiny XII} & ${\rm 1.4754^{+0.0014}_{-0.0031}}$
 & ${\rm 10.24^{+1.82}_{-1.73} }$ & ${\rm 4.09^{+0.99}_{-0.45}}$
 & ${\rm 23.4}$ & ${\rm 0.74^{+0.24}_{-0.11}}$ & ${\rm 0.74^{+0.07}_{-0.04}}$ \\
Ne$\;${\tiny IX} & ${\rm 0.918^{+0.0014}_{-0.0011}}$
 & ${\rm 5.78^{+1.22}_{-0.97} }$ & ${\rm 13.26^{+2.45}_{-2.04}}$
 & ${\rm 23.2}$ & ... & ... \\
Ne$\;${\tiny X} & ${\rm 1.0194^{+0.0016}_{-0.0014}}$
 & ${\rm 7.89^{+1.48}_{-1.34} }$ & ${\rm 12.04^{+1.57}_{-1.86}}$
 & ${\rm 25.8}$ & ${\rm 0.91^{+0.18}_{-0.22}}$ & ${\rm 0.49^{+0.01}_{-0.05}}$ \\
O$\;${\tiny VII} & ${\rm 0.5709^{+0.0017}_{-0.0006}}$
 & ${\rm 4.42^{+0.96}_{-0.81} }$ & ${\rm 87.62^{+11.02}_{-20.95}}$
 & ${\rm 97.5}$ & ... & ... \\
O$\;${\tiny VIII} & ${\rm 0.6530^{+0.0007}_{-0.0008}}$
 & ${\rm 3.90^{+0.68}_{-0.76} }$ & ${\rm 47.15^{+9.23}_{-5.70}}$
 & ${\rm 65.7}$ & ${\rm 0.54^{+0.17}_{-0.09}}$ & ${\rm
 0.22^{+0.02}_{-0.02}}$ 
\enddata
\end{deluxetable}
%
%
\begin{deluxetable}{lccccc}
\tablewidth{0pt}
\tablecaption{Best-fit parameters of a Gaussian to the outburst
 H-like {\ka} lines \label{table:singlegauss}}
\tablehead{ 
&\colhead{$E_{\rm c}$\tablenotemark{a}} & \colhead{$\sigma$} & \colhead{$I$} &
\colhead{$\sigma/E_{\rm c}$} & \colhead{$\chi^2/{\rm d.o.f.}$} \\
\colhead{Ion} & \colhead{[keV]} & \colhead{[eV]}
 & \colhead{[$10^{-4}$ cm$^{-2}$s$^{-1}$]} & [$10^{-3}$]
} 
\startdata 
Si$\;${\tiny XIV} & 2.0052
 & ${\rm 14.34^{+1.48}_{-1.64}}$ & ${\rm 3.74^{+0.41}_{-0.39}}$
 & ${7.15^{+0.74}_{-0.82}}$ & 31.6/35 \\
Mg$\;${\tiny XII} & 1.4724
 & ${\rm 11.13^{+1.92}_{-2.01}}$ & ${\rm 4.39^{+0.70}_{-0.66}}$
 & ${ 7.56^{+1.30}_{-1.37}}$ & 19.7/29 \\
Ne$\;${\tiny X} & 1.0218
 & ${\rm 7.52^{+1.69}_{-1.50}}$ & ${\rm 11.75^{+1.79}_{-1.93}}$
 & ${7.35^{+1.65}_{-1.47}}$ & 21.0/18 \\
O$\;${\tiny VIII} & 0.6535
 & ${\rm 3.48^{+1.06}_{-0.73}}$ & ${\rm 48.37^{+8.01}_{-7.21}}$
 & ${ 5.33^{+1.62}_{-1.12}}$ & 6.1/7
\enddata
\tablenotetext{a}{~Fixed at the laboratory values.}
\end{deluxetable}
%
%
\begin{deluxetable}{lccccc}
\tablewidth{0pt}
\tablecaption{Best-fit parameters of the stratified shell model to the
 outburst H-like {\ka} lines \label{table:trapez}}
\tablehead{
 & \colhead{$E_{\rm c}$\tablenotemark{a}}
 & \colhead{$\theta_{\rm o}$\tablenotemark{b}}
 & \colhead{$r_{\rm sh}/R_{\rm WD}$\tablenotemark{c}}
 & \colhead{$v$\tablenotemark{d}} & \colhead{$\chi ^{2}$/d.o.f.} \\
\colhead{Ion} & \colhead{[keV]} & \colhead{[deg]}
 & \colhead{} & \colhead{[km s$^{-1}$]} & \colhead{}
}
\startdata
Si$\;${\tiny XIV} & 2.0052  & ${\rm 65^{+8}_{-12}}$
 & ${\rm 2.1^{+\infty}_{-0.6}}$ & ${\rm 3920^{+360}_{-280}}$ & 15.5/33\\
Mg$\;${\tiny XII} & 1.4724  & ${\rm 59^{+11}_{-37}}$
 & ${\rm 2.6^{+\infty}_{-1.1}}$ & ${\rm 4030^{+840}_{-470}}$ & 19.0/27\\
Ne$\;${\tiny X} & 1.0218  & ${\rm 44^{+13}_{-16}}$
 & ${\rm 1.3^{+0.4}_{-0.1}}$  & ${\rm 4510^{+1380}_{-530}}$ & 16.9/16\\
O$\;${\tiny VIII} & 0.6535  & ${\rm 42^{+16}_{-17}}$
 & ${\rm 1.3^{+0.6}_{-0.2}}$  & ${\rm 3380^{+780}_{-550}}$ & 6.1/5
\enddata
\tablecomments{~Inclination is assumed to be 40$^{\rm \circ}$.}
\tablenotetext{a}{~Fixed at the laboratory values.}
\tablenotetext{b}{~Opening polar angle of the shell within which the plasma is absent.}
\tablenotetext{c}{~Ratio of the radius of the shell to that of white dwarf.}
\tablenotetext{d}{~Radial velocity of the shell.}
\end{deluxetable}

\begin{deluxetable}{lccccc}
\tablewidth{0pt} 
\tablecaption{Best-fit parameters of the {\sc diskline} model to the
 outburst H-like {\ka} lines \label{table:diskline}} 
\tablehead{ 
 & \colhead{$E_{\rm c}$\tablenotemark{a}}
 & \colhead{$q$\tablenotemark{b}}
 & \colhead{$R_{\rm in}$} & \colhead{$I$}
 & \colhead{$\chi^{2}$/d.o.f.} \\
\colhead{Ion}    & \colhead{[keV]}       &
 & \colhead{[$10^{8}$ cm]}
 & \colhead{[$10^{-4}$ cm$^{-2}$s$^{-1}$]} & \colhead{}
}
\startdata
Si$\;${\tiny XIV} & 2.0052
 & ${\rm -3.31^{+0.44}_{-0.74}}$ & ${\rm 10.4^{+2.2}_{-1.9}}$
 & ${\rm 3.73\pm 0.39}$ & 15.4/34\\
Mg$\;${\tiny XII} & 1.4724
 & ${\rm -2.73^{+0.36}_{-0.55}}$ & ${\rm 7.0^{+3.5}_{-2.4}}$
 & ${\rm 4.26^{+0.71}_{-0.66}}$ & 17.4/28 \\
Ne$\;${\tiny X} & 1.0218
 & ${\rm -2.15^{+0.18}_{-0.23}}$ & ${\rm 2.1^{+1.4}_{-0.9}}$
 & ${\rm 12.55^{+2.04}_{-2.13}}$ & 19.6/17 \\
O$\;${\tiny VIII}  & 0.6535
 & ${\rm -1.98^{+0.22}_{-0.23}}$ & ${\rm 2.5^{+4.7}_{-2.2}}$
 & ${\rm 53.60^{+4.56}_{-8.29}}$ & 2.3/6
\enddata
\tablenotetext{a}{~Fixed at the laboratory values.}
\tablenotetext{b}{~Radial power of the line emissivity
 $\varepsilon \propto r^q$.}
\end{deluxetable}

\begin{deluxetable}{lcccccc}
\tablewidth{0pt}
\tablecaption{Best-fit parameters of the double Gaussian model to the
 outburst H-like {\ka} lines \label{table:doublegauss}}
\tablehead{
 & \colhead{$E_{\rm c}$\tablenotemark{a}}
 & \colhead{$\sigma\tablenotemark{b}/E_{\rm c}$}
 & \colhead{$E_{\rm sp}\tablenotemark{c}/E_{\rm c}$}
 & $v\sin i\tablenotemark{d}$ & \colhead{$\chi ^{2}$/d.o.f.}\\
\colhead{Ion} & \colhead{[keV]} & \colhead{[${\rm \times 10^{-3}}$]} & 
\colhead{[${\rm \times 10^{-3}}$]} & [km/sec]
}
\startdata
Si$\;${\tiny XIV} & 2.0052
 & ${\rm 3.8^{+0.8}_{-0.5}}$ & ${\rm 10.6\pm 1.1}$
 & ${\rm 1590\pm 170}$ & 19.4/34 \\
Mg$\;${\tiny XII} & 1.4724
 & ${\rm 4.6^{+3.8}_{-1.0}}$ & ${\rm 9.9^{+2.3}_{-2.9}}$
 & ${\rm 1490^{+350}_{-430}}$ & 18.0/28 \\
Ne$\;${\tiny X} & 1.0218
 & ${\rm 7.6^{+1.5}_{-3.2}}$ & $<8.6$
 & $<1290$ & 21.1/17 \\
O$\;${\tiny VIII}  & 0.6535
 & ${\rm 5.3^{+1.6}_{-2.6}}$ & $<7.3$
 & $<1100$  & 6.1/6
\enddata
\tablenotetext{a}{~Fixed at the laboratory values.}
\tablenotetext{b}{~Common width of the blue- and redshifted components in
 Gaussian $\sigma$.}
\tablenotetext{c}{~Energy separation of the blue- and redshifted components.}
\tablenotetext{d}{~Average line-of-sight velocity of the blue- or redshifted
 line component calculated from $E_{\rm sp}/E_{\rm c}$.}
\end{deluxetable}

%
%
%
%

\end{document}